\documentstyle[aps,a4]{revtex}

\newcommand{\bra}[1]{\langle #1 |\,}

\newcommand{\ket}[1]{\,| #1 \rangle}
\newcommand{\Ket}[1]{\left| #1 \right\rangle}
\newcommand{\bracket}[2]{\langle #1 | #2 \rangle}

\newcommand{\G}{G}
\newcommand{\mat}[1]{\underline{\underline{#1}}}
\newcommand{\mG}{\mat{G}}

\newcommand{\cH}{{\hat{{\cal{H}}}}}
\newcommand{\cG}{{\cal{G}}}

\newcommand{\m}{\hat{\mu}}

\newcommand{\bpsi}{\bra{\Psi_o}}
\newcommand{\kpsi}{\ket{\Psi_o}}

\newcommand{\dv}[2]{\left(\begin{array}{c} #1 \\ #2 \end{array}\right)}
\newcommand{\tv}[3]{\left(\begin{array}{c} #1 \\ #2 \\ #3 \end{array}\right)}

\newcommand{\id}{\mbox{\bf id}}
\newcommand{\kv}{\ket{\mbox{vac}}}
\newcommand{\bv}{\bra{\mbox{vac}}}
\newcommand{\Self}{{\cal{S}}}
\newcommand{\mS}{\mat{\Self}}

\newcommand{\beq}{\begin{equation}}
\newcommand{\eeq}{\end{equation}}
\newcommand{\beqs}{\begin{eqnarray}}
\newcommand{\eeqs}{\end{eqnarray}}
\newcommand{\zero}{{(0)}}		
\newcommand{\Ho}{{\hat{H}_o}}	

\newcommand{\papa}{particle--particle }

\newcommand{\Y}{{\mbox{\sf Y}}}		
\newcommand{\Eo}{{E_o^N}}
\newcommand{\Hp}{{\hat{H}_{p}}}		
\newcommand{\projectile}{{({p})}}		
\newcommand{\ham}{{\hat{H}}}		
\newcommand{\K}{{{K}}}
\newcommand{\calH}{{\cal H}}
\newcommand{\p}{{p}}
\newcommand{\q}{{q}}
\newcommand{\ik}{{k}}
\newcommand{\E}{{\cal E}}
\newcommand{\opg}{single--particle Green's function}
\newcommand{\ky}[1]{| Y_{#1} \rangle}
\newcommand{\by}[1]{\langle Y_{#1} |}
\newcommand{\ddrs}{\delta_{rr'}\delta_{ss'}- \delta_{rs'}\delta_{sr'}}
\newcommand{\mcG}{{\mat{\cG}}}
\newcommand{\cGs}{{\tilde{\cG}}}	

\newcommand{\Selfs}{{\tilde{\Self}}}	
\newcommand{\cTs}{{\tilde{\cal T}}}	
\newcommand{\cT}{\cTs}
\newcommand{\scat}[2]{S({#1}\leftarrow{#2})}
\newcommand{\scatT}[2]{t({#1}\leftarrow{#2})}

\title{Extended Two--Particle Green's Functions and Optical Potentials for 
	Two--Particle Scattering by Many--Body Targets }

\author{Joachim Brand and Lorenz S. Cederbaum\\
	Theoretische Chemie, Universit\"at Heidelberg, \\
	Im Neuenheimer Feld 253, D--69120 Heidelberg}

\begin{document}

\maketitle

\begin{abstract}
An extension 
of the fermionic \papa propagator 
is presented, that possesses similar 
algebraic properties to the \opg. In particular, this extended two--particle Green's function satisfies
Dyson's equation and its self energy has the same analytic structure as the the self energy
of the \opg. For the case of a system interacting by one--particle potentials only, the two--particle
self energy takes on a particularly simple form, just like the common self energy does.
The new two--particle self energy also serves as a well behaved optical potential for the elastic
scattering of a two--particle projectile by a many--body target. Due to its analytic structure, the
two--particle self energy avoids divergences that appear with effective potentials derived by other
means.
\end{abstract}

\vspace{1cm}
{\sf preprint submitted 1996 to  } {\em Annals of Physics (N. Y.)}


\setcounter{section}{0}
\setcounter{equation}{0}
\setcounter{footnote}{0}

\section{Introduction} \label{s1}

Many--body Green's functions, also called propagators, provide useful tools for investigating
various properties of many--body systems like solids, atomic nuclei, atoms, and molecules.
The so called \opg\cite{fetter} is the most fundamental in a whole hierarchy of related propagators.
One of the reasons for the outstanding importance of the \opg\ is that it satisfies a well--known 
equation --- the Dyson equation\cite{fetter}. Among the wealth of applications of this
equation we want to mention that the Dyson equation has been exploited for calculations of
ionisation spectra of molecules\cite{Cederbaum:86,Cederbaum:90,Zakrewski:95} as well as scattering
cross sections\cite{dieterN2Pig}. In fact, the Dyson equation allows one to derive exact effective 
one--particle equations for the elastic scattering of a particle by a many--body 
target\cite{bellandsquires,namiki}. In such a one--particle equation, a nonlocal and in general
energy--dependent potential --- called {\em optical potential} --- includes the correlations of the
many--particle system.
The Green's function optical potential --- also called {\em self energy} --- has several advantages
upon the optical potentials derived by other means, 
e.~g.~by Feshbach projection\cite{feshbach_I,feshbach_II} from the multiple particle 
Schr\"odinger equation.
One point is the good--natured energy dependence that avoids  divergences for high energies.
Also the self energy closely resembles the properties of the phenomenological optical model potentials
that have been common in nuclear physics for a long time\cite{CapuzziMahaux}.
Further we want to mention that the self energy of a system interacting by one--particle potentials
takes on a very simple appearance and provides an exact optical potential already 
in first order perturbation theory. This is usually not the case for optical potentials derived
by other means.
In general, the well developed approximation schemes
for the self energy based on perturbation theory\cite{adc_jochen}
account for a balanced treatment of the correlation, both in the
target and the scattering state. This balanced treatment is much more difficult to achieve in
wavefunction approaches\cite{dieterN2Pig}.

The next member in the hierarchy of Green's functions is the two--particle Green's function\cite{fetter}.
This is a function of three energy variables.
Although this function satisfies an integral equation with kernels that can be approximated
in a diagrammatic approach --- the Bethe--Salpeter equation --- 
for most problems it is too complicated to handle. Instead, one uses
propagators with only one energy variable left, e.~g.~the {\em polarization
propagator}\cite{fetter,pines} and
the {\em \papa propagator}\cite{gross_runge} which have 
been used to calculate excitation energies\cite{oddershede_rev,sascha} 
and double--ionisation spectra\cite{FTarantelli:94,FTarantelli:93,Argen:92}, respectively.
Furthermore, approaches have been reported starting from the latter propagator to calculate amplitudes for two particle
scattering via optical potentials\cite{junkin_I,junkin_II}.
Although there are some analogies between these propagators and the \opg, the Dyson equation is
not passed on to these two propagators. Neither by diagrammatical analysis nor from
the algebraic point of view is there any close analogon to this fundamental equation.

The main goal of this paper is to show that it is possible to construct extensions of these
two--particle propagators (only the \papa propagator for fermions is considered explicitly) that
have the major algebraic properties in common with the \opg, namely those that lead to
a Dyson equation and a {\em two--particle self energy}. 
It is particularly interesting to note
that this two--particle self energy serves as an {\em optical potential}
for  the elastic scattering of a  two--fermion projectile
by a correlated target. 
Possible applications include, e.~g., the scattering of a Deuteron by an atomic nucleus or the 
scattering of a Positronium by a molecule.
In general, the usefulness and quality of possible
approximations are determined by the algebraic properties of the optical potential.
The two outstanding features of our new, two--particle self energy are
the good--natured energy dependence that has a finite limit for high energies and the
fact that influences resulting 
from a one--particle potential are accounted for exactly already in the energy independent 
first order. These properties are straightforward consequences of the
algebraic analogies to the \opg, but they are in no way a matter of course with two--body
effective potentials.

This paper is organized as follows: After briefly reviewing the algebraic properties of the
\opg\ we will show in section \ref{s3} by explicit construction, that an extension of the \papa 
propagator
is possible that inherits the major algebraic properties from the \opg\ and satisfies a Dyson equation.
Discussing the properties of the extended propagator and its self energy we also
consider the special case of a system interacting only by one--particle potentials.
In section \ref{s2} we then  show how the extended propagator can be used for deriving
well behaved effective two--particle scattering equations. We will, however, develop these ideas
in a general framework allowing for either bound or free two--particle projectile states.
Also, we will state explicitly the properties of the extended two--particle propagator that
are  prerequisites for derivation of effective scattering equations.

\section{Construction of Two--Particle Propagators} \label{s3}

In this section
we will construct an extended two--particle propagator 
that inherits a couple of fundamental properties from the
\opg. Although the \opg\ and
the Dyson equation are well known and described thoroughly in many standard
textbooks\cite{fetter,gross_runge}, we briefly review the algebraic 
derivation of the Dyson equation\cite{alessandra}, in order to motivate our construction
of extended two--particle propagators. We carry on by applying the formalism
developed for the \opg\ to  the \papa propagator and construct one possible
example for an extension of it. This extension has similar algebraic properties to the
\opg. We show that it is possible to derive Dyson's equation analogously and
define a two--particle self energy.
We examine the properties in relation to the \opg.
In particular we consider the case of
a system interacting with a one particle potential only. 
In order to be able to apply the
extended propagator to the  scattering problem of section \ref{scat_sect}, we
Fourier transform the propagator and Dyson's equation into time space. 

\subsection{Single--Particle Green's Function and Dyson Equation} \label{section_opg}

The {\em \opg}\cite{fetter} in energy space is defined by:
\beqs \label{def_opg}
	\nonumber
	G_{pq}(\omega) &=& \bpsi\, a_p \frac{1}{\omega-\ham+\Eo+i\eta} a^\dagger_q\, \kpsi\\
	& & + \bpsi\, a^\dagger_q \frac{1}{\omega-\Eo+\ham-i\eta} a_p\, \kpsi
\eeqs
Here, $\kpsi$ denotes the bound ground state of the correlated $N$--fermion system
with Energy $\Eo$. 
The operator $a_q^\dagger$ (or $a_q$) creates (destroys) one fermion
in an orbital that belongs to the one--particle quantum number $q$. The set of possible
indices $\{q\}$ covers the one fermion Hilbert space.
The positive infinitesimal $\eta$
provides the correct time ordering when the expressions are Fourier transformed into time space.
The Hamiltonian $\ham$ is the usual Fock space Hamiltonian that contains the kinetic energy
as well as possible interactions with external potentials and two body interactions.
As usual it is convenient to present it as a sum
of a {\em zeroth order} part $\Ho$ and an {\em interaction} part $\ham_1$: 
\beq
	\ham = \Ho + \ham_1
\eeq
We choose the zeroth
order Hamiltonian to be a diagonal 
one particle operator with respect to the underlying orbital basis:
\beq \label{defHo}
	\Ho = \sum_i \varepsilon_i\, a_i^\dagger a_i
\eeq

The two parts of the \opg\ describe the propagation of quasi--particles and quasi--holes
in the correlated $N$--fermion system.
Formally we can interpret each of the two parts as  a projection of an operator
resolvent onto the states $a_p\,\kpsi$ and $a^\dagger_q\, \kpsi$ respectively. These
states belong to the Hilbert spaces of $N\pm 1$ fermions. They are neither orthogonal
among each other nor normalized, since the scalar products yield one particle
density matrices:
\beq
	\nonumber
	\bpsi\, a_p  a^\dagger_q\, \kpsi\quad\mbox{or}\quad\bpsi\, a^\dagger_q  a_p\, \kpsi
\eeq
If, however, we formally introduce the  {\em composite states}
\begin{equation} \label{def_YG}
	\ket{Y_q^\G} := \dv{ a^\dagger_q \kpsi}{\bpsi a^\dagger_q},
\end{equation}
these form an orthonormal set with respect to a suitably defined scalar product:\footnotemark
\begin{eqnarray}
	\nonumber \bracket{Y_p^\G}{Y_q^\G}
	&=& \bpsi a_p  a_q^\dagger \kpsi + \bpsi a_q^\dagger   a_p  \kpsi \\
	&=& \bpsi \{ a_p, a_q^\dagger\} \kpsi =
	\delta_{pq} 
\end{eqnarray}
Mathematically speaking the composite states live in the direct sum of the $N+1$ particle
Hilbert space and the dual of the $N-1$ particle space.

The states $\ket{Y_q^\G}$ allow the following shorthand notation for the \opg:
\beq \label{G_by_Y}
	\G_{pq}(\omega) = \bra{Y_p^\G} \frac{1}{\omega - \cH^\G} \ket{Y_q^\G}
\eeq
The operator $\cH^\G$ in the composite space is defined by:
\beq\label{def_cHG}
	\cH^\G := \left(\begin{array}{cc}
	\hat{H} - E_o^N - i \eta	&	0	\\
	0			& E_o^N - \hat{H} + i\eta \\
	\end{array} \right) 
\eeq
Eq.~(\ref{G_by_Y}) now  allows us to interpret the full \opg\ as the projection of an operator
resolvent onto a set of orthonormal states.

These states $\{\ket{Y_q^\G}\}$ span a subspace of the extended Hilbert space they live in.
We want to draw attention to the point, that
these states are labelled by the complete set of one particle quantum numbers, regardless, whether
or not the indices refer to 
particle or hole states. Due to this fact, the spanned subspace 
has the same dimension as (i.~e.~is isomorphous to)
the one particle Hilbert space. Now we can extend the set $\{\ket{Y_q^\G}\}$ to an orthonormal
basis $\{\ket{Q_J^\G}\}$ of the composite space and use  matrix partitioning\footnotemark
\ in order
to derive an expression for the
inverse of the matrix $\mG(\omega)$ with elements given by
Eq.~(\ref{G_by_Y}).
This directly yields
{\em Dyson's equation}. We present it in the following form:
\begin{equation} \label{dyseq}
	{\mG(\omega)}^{-1} =  \omega \mat{1} - \mat{\varepsilon} - \mat{\Sigma}(\omega)
\end{equation}
Here the matrix $\mat{\varepsilon}$ is diagonal and contains the zeroth order single--particle energies
$\varepsilon_p$ as well as the infinitesimals $\pm i\eta$.
The {\em self energy} $\mat{\Sigma}(\omega)$ is defined by
\begin{equation} \label{self_energy}
	\mat{\Sigma}(\omega) =  \mat{\calH}_{aa} -\mat{\varepsilon} +
	\mat{\calH}_{ab}
	\frac{1}{\omega - \mat{\calH}_{bb}}\mat{\calH}_{ba}.
\end{equation}
Here $\mat{\calH}$ is a matrix representation of the operator $\cH$
in the basis $\{\ket{Q_J^\G}\}\supset\{\ket{Y_q^\G}\}$ of the extended space:
\begin{eqnarray} 
	\left[\mat{\calH}\right]_{IJ} &=& \bra{Q_I^\G}\cH^\G\ket{Q_J^\G}
\end{eqnarray}
By virtue of the subdivision of the basis set into two parts, the matrix is structured into blocks:
\begin{eqnarray} \label{block_H}
	{\mat{\calH}}& =& \left(\begin{array}{cc}
	\mat{\calH}_{aa}	&	\mat{\calH}_{ab}	\\
	\mat{\calH}_{ba}	&	\mat{\calH}_{bb}
	\end{array} \right) 
\end{eqnarray}
The block index $a$ refers to the set $\{\ket{Y_q^\G}\}$, the index $b$ to the complemental part of the
basis.
The zeroth order of $\mat{\calH}_{aa}$ is given by $\mat{\varepsilon}$.
For more details on the theory we refer to  \cite{alessandra}.

It depends on the point of view, of course, whether the Dyson equation (\ref{dyseq}) is seen as
an equation that relates quantities which are defined separately or as
a definition of the self energy.
In the latter case, it is Eq.~(\ref{self_energy}) that has to be derived. We want to remind the
reader, that historically the self energy was defined through its perturbation series and consequently
the Dyson equation had to be proven\cite{dyson1,dyson2}. 
In contrast to this historic background we understand the Dyson
equation rather as an algebraic property of the \opg . When we say that we try to construct two--particle 
propagators that satisfy a Dyson equation, we mean it exactly in this sense.

\subsection{Two--Particle Propagators} \label{twoparticleprops_section}

After this excursion to the \opg\ we will focus our attention on two--particle propagators.
We consider the \papa propagator\cite{gross_runge}
and its constituents:
\beqs \label{def_papaprop}
	\Pi_{rs,r's'}(\omega) &=& \Pi^+_{rs,r's'}(\omega) - \Pi^-_{rs,r's'}(\omega) \\
\label{def_papa_plus}
	\Pi^+_{rs,r's'}(\omega) &=& \bpsi a_r a_s \frac{1}{\omega-\ham+\Eo+i\eta}
		 a^\dagger_{s'} a^\dagger_{r'} \kpsi \\\label{def_papa_minus}
	\Pi^-_{rs,r's'}(\omega) &=& \bpsi a^\dagger_{s'} a^\dagger_{r'}\frac{1}{\omega-\Eo+\ham-
i\eta}
		 a_r a_s  \kpsi
\eeqs
The part $\Pi^+_{rs,r's'}(\omega)$ contains the relevant information for elastic two particle
scattering off the correlated ground state,  whereas $\Pi^-_{rs,r's'}(\omega)$ can be used to
calculate double ionization spectra.
We would like to mention that when transformed to time--space, the \papa propagator is just an
additive part of the full two--particle Green's function consisting of two specific
time--orders of the latter.

The analogy to the \opg\ fails in the point that neither the $N\pm 2$ particle states
\beq
	\nonumber
	a^\dagger_{s} a^\dagger_{r} \kpsi\quad\mbox{or}\quad a_r a_s  \kpsi
\eeq
nor the composite vectors
\beq
	\nonumber
	\dv{ a^\dagger_{s} a^\dagger_{r} \kpsi}{\bpsi a^\dagger_{s} a^\dagger_{r} }
\eeq
satisfy any kind of orthogonality relation.
Even more, the overlap matrices do not have to be regular at all.
This is most easily seen by considering the special case of a noninteracting system (zeroth order).
In this case, the states defined above are slater determinants. All these states, however, vanish, if
the index $r$ refers to an orbital that is occupied in the ground state determinant and $s$
refers to an unoccupied orbital. I.~e.~,the space spanned by these states is not isomorphous
to the two particle Hilbert space but has a lower dimension.

Comparing with the one particle case, we see, that likewise the states $a^\dagger_q \kpsi$
do not span an isomorphous space to the one particle space, but the composite states $\ket{Y_q^\G}$ 
do. That was the reason for being able to express the \opg\ as a {\em matrix inverse} or resolvent
in a space spanned
by the complete set of one--particle quantum numbers.
The fact that the states $\ket{Y_q^\G}$ form an orthonormal set has the consequence that
the self energy does not contain any  linear contributions in the frequency variable $\omega$.

In order to restore the analogy to the \opg\ we introduce  the following extended composite states:
\begin{equation} \label{def_ky}
	\ky{rs}=\tv{ a^\dagger_{s} a^\dagger_{r} \kpsi}{\chi_{rs}}{\bpsi a^\dagger_{s} a^\dagger_{r} }
\end{equation}
With the choice
\begin{equation} \label{def_Chim}
	\chi_{rs}:= a^\dagger_{s} \kv \otimes \bpsi a^\dagger_{r} - a^\dagger_{r} 
	\kv \otimes \bpsi a^\dagger_{s},
\end{equation}
where $\kv$ denotes the Fock--space state with no particles, and the ``metric''
\beq \label{choosemetric}
	\m :=\left(\begin{array}{ccc}
	1 & 0 & 0 	\\
	0 & 1 & 0 	\\ 
	0 & 0 & -1 	\\
	\end{array} \right)
\eeq
these states fulfil the following ``orthogonality'' relation:\footnotemark
\beq \label{orthogonal}
	\by{rs}\,\m\,\ky{r's'} = \delta_{rr'}\delta_{ss'}- \delta_{rs'}\delta_{sr'}
\eeq

This is as much as a unit matrix in a space spanned by two particle quantum numbers. The second term
on the right hand side of Eq.~(\ref{orthogonal}) is due to antisymmetry in the one particle indices
that comes from the fermionic nature of the particles.

The extended states live in a composite vector space, denoted by $\Y$, which is the direct sum of
the spaces, the three components live in. These are: the $N+2$ particle space, the dual of the 
$N-2$ particle space, and the space $\chi_{rs}$ lives in, which is the tensor product of the
one particle Hilbert space and the dual of the $N-1$ particle space.
We have found other possible choices for the extension $\chi_{rs}$ that live in  different
spaces and utilize different metrics\cite{diplom}. In this paper we only give one example.

In the composite space $\Y$, the metric $\m$ defines a binary product by:
\beq
	(\ket{A}\,,\,\ket{B}) = \bra{A}\,\m\,\ket{B}
\eeq
We want to mention that due to the indefinite choice of the metric $\m$, this binary product
is not a scalar product.

In order to define a Green's function analogous to Eq.~(\ref{G_by_Y}), we first 
have to define an operator that takes on the role of $\cH^\G$ from Eq.~(\ref{def_cHG}).
A suitable choice for our goals is:
\beq \label{def_cH}
	\cH := \left(\begin{array}{ccc}
	\hat{H} - E_o^N - i \eta	&	0	&	0	\\
	0	&	\hat{H}^{(1)}+\Eo - \hat{H}^{(N-1)} + i\eta	&	0	\\
	0	&	0	& E_o^N - \hat{H} + i\eta \\
	\end{array} \right) 
\eeq
The symbol $\hat{H}^{(1)}$ denotes an operator that acts like the usual Fock space Hamiltonian
on the first component (the one particle part) 
of the product space, i.~e.~$\hat{H}^{(1)} \equiv \hat{H}\otimes\hat{1}$.
Accordingly, $\hat{H}^{(N-1)}$ acts only on the second component, which is the dual of the
$N-1$ particle space, i.~e.~$\hat{H}^{(N-1)} \equiv \hat{1}\otimes\hat{H}$.

Now we are in the position that we can define a new {\em extended two--particle Green's function}
by
\beqs \label{def_cG}
	\cG_{rs,r's'}(\omega)& :=& \by{rs\,}\m\frac{1}{\omega - \cH}\,\ky{r's'}
\eeqs
It follows directly from this definition that this function contains the \papa propagator
as an additive component:
\beqs \label{cG}
	\nonumber\lefteqn{
	\cG_{rs,r's'}(\omega) = \Pi_{rs,r's'}(\omega)} \\
	& & + \left\langle \chi_{rs}\,,\,
	\frac{1}{\omega - \hat{H}^{(1)}-\Eo + \hat{H}^{(N-1)} - i\eta}\,\chi_{r's'}\right\rangle\\
	\nonumber 
	&=& \Pi_{rs,r's'}(\omega) \\
	\nonumber
	& & +\left\{\frac{1}{i} \bv \,a_s\frac{1}{\omega - \hat{H} -i\eta} a_{s'}^\dagger\, \kv
	\ast  \bpsi\, a^\dagger_{r'} \frac{1}{\omega-\Eo+\ham-i\eta} a_r\, \kpsi\right\} \\
	\label{cG_ast}
	& &- \{ r \longleftrightarrow s \}- \{ r' \longleftrightarrow s' \}
	+ \{ r \longleftrightarrow s\,,\,r' \longleftrightarrow s' \}
\eeqs
The asterix $\ast$ denotes convolution with respect to $\omega$ and the symbol 
$\{ r \longleftrightarrow s \}$ stands for the term  in braces reappearing with interchanged
indices.

The new terms that appear in the extended propagator additionally to the \papa propagator
are analytic functions of $\omega$ in the lower half of the complex plane. Right above
the real axis they feature single poles and branch cuts. The poles appear at energies that are
sums of single--particle energies (eigenvalues of $\hat{H}^{(1)}$) and quasi--hole energies
(single ionization energies or eigenvalues of $\hat{H}^{(N-1)}$). Branch cuts originate in the
continuous parts of the spectra of the single--particle Hamiltonian $\hat{H}^{(1)}$ and
the $(N-1)$--particle Hamiltonian $\hat{H}^{(N-1)}$, respectively.
In computations with finite basis sets, of course, only discrete single poles appear.
In general, the new poles introduced in the extension $\mat{\cG}(\omega)$ of the \papa
propagator $\mat{\Pi}(\omega)$ are distinct from the poles of the latter.
Therefore, the quantities of interest, namely the poles and residues of the \papa propagator,
can be identified and extracted from the extended propagator.

\subsection{Dyson Equation for the Extended Propagator}

The starting point for our considerations is definition (\ref{def_cG}) and the orthogonality
relation (\ref{orthogonal}) for the composite states $\ky{rs}$.
Since these states are antisymmetric with respect to a permutation of the indices $r$ and $s$,
we introduce the index restriction $r>s$ in order to make the set of vectors $\{\ky{rs}\}$
linear independent. Now we can extend this set to a basis $\{\ket{Q_I}\}\supset\{\ky{rs}\}$
of the space $\Y$. We may choose this set orthogonal with respect to the binary product:
\beq \label{orth_Q}
	\bra{Q_I}\,\m\,\ket{Q_J} = \delta_{IJ}\,m_I
\eeq
Due to the indefiniteness of the metric $\m$, we cannot normalize all states to
$m_I = +1$, but we can require that
\beq
	m_I \in \{-1, +1\}
\eeq
We define the following matrices:
\beqs
	\left[\mat{\mu}\right]_{IJ} &:=& \delta_{IJ}\,m_I = \bra{Q_I}\,\m\,\ket{Q_J} \\
	\Big[\mat{\calH}\Big]_{IJ} &:=& \bra{Q_I}\,\m\cH\,\ket{Q_J}
\eeqs
Just like in section \ref{section_opg}, Eq.~(\ref{block_H}), 
these matrices are structured into blocks by the
subdivision of the basis $\{\ket{Q_I}\}$ into the subset $\{\ky{rs}\}$ and its complement.
Of course, the upper left block of $\mat{\mu}$ is just the unit matrix due to equation
(\ref{orthogonal}):
\beq
	\mat{\mu}_{aa} = \mat{1}
\eeq

After some simple manipulations, we can express our extended propagator as the upper
left corner of an inverse matrix:
\beq
	\cG_{rs,r's'}(\omega) = \left[\frac{1}{\omega\mat{\mu} - \mat{\calH}}\right]_{rs,r's'}
\eeq
By simple matrix partitioning we now find:
\beqs \label{dys1}
	\left[\mcG(\omega)\right]^{-1} = \omega\mat{1} - \mat{\calH}_{aa} -
	\mat{\calH}_{ab}
	\frac{1}{\omega\mat{\mu}_{bb} - \mat{\calH}_{bb}}\mat{\calH}_{ba}
\eeqs
This equation already reveals the algebraic structure we were aiming at. 
Of course, we can make it look more like Dyson's
equation by introducing the matrix
\beqs
	\Big[\mat{\varepsilon}\Big]_{rs,r's'} := (\ddrs)(\varepsilon_r + \varepsilon_s)
\eeqs
which is easily recognized as the  zeroth order of $\mat{\calH}_{aa}$ apart from infinitesimals
$\cal{O}(\eta)$. When we introduce
the ``self energy'' for the extended propagator by
\beq\label{dys2_self}
	\mS(\omega) = \mat{\calH}_{aa} -\mat{\varepsilon} +
	\mat{\calH}_{ab}
	\frac{1}{\omega\mat{\mu}_{bb} - \mat{\calH}_{bb}}\mat{\calH}_{ba}
\eeq
Eq.~(\ref{dys1}) becomes {\em Dyson's equation for the extended two--particle propagator:}
\beqs\label{dys2}
	\Bigg[\mcG(\omega)\Bigg]^{-1} = \omega\mat{1} -\mat{\varepsilon} -\mS(\omega)
\eeqs 
Obviously, the zeroth order of the extended propagator takes on the following appearance:
\beq \label{resolvent_zeroth}
	\left[\mcG^\zero(\omega)\right]^{-1} = \omega\mat{1} - \mat{\varepsilon}
\eeq
where the infinitesimal $i\eta(2\overline{n}_r\overline{n}_s-1)$ has been left out for
simplicity. The anti--occupation number $\overline{n}_r$ takes on the value 1 if the index $r$
refers to an orbital unoccupied in the zeroth order ground state (virtual orbital) and
0 if otherwise.

Finally we can express the Dyson equation (\ref{dys2}) in the  appealing equivalent form:
\beq\label{dys3}
	\mcG(\omega) = \mcG^\zero(\omega) + \mcG^\zero(\omega)\mS(\omega)\mcG(\omega)
\eeq

We note that the self energy $\mS(\omega)$ can be calculated from Eq.~(\ref{dys2_self}).
Of course, the choice of the extended basis is not unique. Different ways of constructing
finite subsets of this basis will hence lead to  different approximations for the self energy.
For example, powerful 
approximations may be achieved by constructing the basis set from {\em correlated
excited states}, classified by the nature of the excitation  with respect to the ground state.
This is a common technique used in many modern many--body methods\cite{frank_I,frank_II}.

By construction, the extended Green's function $\mcG(\omega)$ contains the full \papa 
propagator and, therefore, it can be used not only for scattering but also for calculating
{\em double ionization energies} and the corresponding transition amplitudes. One has to
be careful, however, when employing perturbation theory to extract
approximations to the self energy
$\mS(\omega)$ for this purpose. The reason is that there are states in the extended
basis contributing to $\mat{\calH}_{bb}$ and degenerate in zeroth order with 
states $\ky{rs}$ where $r$ and $s$ are hole indices (i.~e., assign 
occupied orbitals in the unperturbed ground state).
It is, of course, possible to obtain valid approximations by applying the formalism
of degenerate perturbation analysis.
The deeper reason
for the appearance of these degeneracies lies in the indefinite choice of the metric $\m$
in Eq.~(\ref{choosemetric}). It is an open question, whether suitable extensions
$\chi_{rs}$ in (\ref{def_ky}) can be found that avoid the degeneracies by using a definite 
metric.
For a more detailed discussion of the
properties of $\mcG(\omega)$, of perturbation theory, and of 
different choices of extensions $\chi_{rs}$ see reference
\cite{diplom}.

\subsection{Resolvents and Energy Dependence} \label{sectResolvents}

A striking new property of the extended propagator is revealed when we compare the
various one-- and two--particle propagators and try to interpret them as resolvents
of some operator.
We  remark that, of course, all the matrices in the two--particle quantum numbers
$(r,s)$ are not matrices in a literal sense, since they are in general of infinite dimension
and the indices may be continuous. These matrices can rather be understood as linear operators
in the two particle Hilbert space. The matrix $\mat{\varepsilon}$ is then 
a representation of the zeroth order Hamiltonian $\Ho$,
restricted to the two particle space.
In this sense, the zeroth order propagator $\mcG^\zero(\omega)$ is a resolvent of this 
Hamiltonian, as Eq.~(\ref{resolvent_zeroth}) shows. 
Of course, the full extended propagator $\mcG(\omega)$ can be seen as
a resolvent of the energy--dependent operator
\[
	\mat{\varepsilon} + \mS(\omega)
\]
In section \ref{eigenv_eqn} we will see that indeed this operator is an effective two particle
Hamiltonian that can be used to describe elastic two--particle scattering off a correlated
$N$--fermion target. We want to remind the reader that neither the \papa propagator
$\mat{\Pi}(\omega)$, nor one of its components $\mat{\Pi}^\pm(\omega)$ is a resolvent
in the complete two particle space, like the discussion at the beginning of 
\ref{twoparticleprops_section} shows.
Still, in a restricted space these propagators may be
understood as resolvents. But the energy--dependent effective Hamiltonians that could be derived
this way are pathological, since they contain linear contributions in $\omega$ that diverge
at high energies.
The energy dependence of the
two--particle self energy $\mS(\omega)$, on the other hand, is given 
by Eq.~(\ref{dys2_self}), where the analytic structure is determined through a partial 
fraction series. For practical purposes this implies that the self energy
has a finite limit for values of the energy variable $\omega$ far outside the spectrum
of the Hamiltonian.

\subsection{One--Particle Potentials} \label{sectOneP}

Another  new feature of the extended two--particle propagators  concerns the
behaviour with respect to one--particle potentials. 
For a system interacting by one--particle potentials the Hamiltonian $\hat{H}$ 
is a one--particle operator, i.~e.~it can be written in the form
\beq \label{one_particle_ham}
	\hat{H}= \sum_{ij} h_{ij} \, a^\dagger_i a_j
\eeq
In this case, the particles of the system do not interact by real two--body forces but
only with external or mean field potentials. Any operator of the form (\ref{one_particle_ham})
can be diagonalized (i.~e.~$h_{ij}\propto \delta_{ij}$) by a simple unitary transformation 
$t_{ij}$ of
the set of orbitals (basis of one--particle wavefunctions). In general, this transformation
includes the full one particle space. It can be seen directly from the definitions
(\ref{def_cG}), (\ref{def_opg}), (\ref{def_papaprop}), (\ref{def_papa_plus}), and
(\ref{def_papa_minus}), that
all of the propagators we have introduced so far, transform with one unitary transformation
matrix on each one--particle index.
For example, the \papa propagator transforms like:
\beq
	\tilde{\Pi}_{ij,kl} = \sum_{i'j'k'l'} t^*_{ii'}\, t^*_{jj'}\, {\Pi}_{i'j',k'l'}
	\, t_{k'k}\, t_{l'l} 
\eeq
Since a diagonalized one particle Hamiltonian is equivalent to the zeroth order Hamiltonian 
(i.~e.~it can be written in the form of Eq.~(\ref{defHo})), 
we can easily study the behaviour of the various
propagators under the influence of a one--particle potential by performing the inverse transformation
on the zeroth order propagators. It is particularly interesting to observe what happens under
this procedure to  equations (\ref{resolvent_zeroth}) and (\ref{dys2}) for our extended 
two--particle propagator, and Eq.~(\ref{dyseq}) for the \opg. From the argumentation given
above, it follows immediately that in both cases the self energies $\mS(\omega)$ and 
$\mat{\Sigma}(\omega)$ become {\em energy independent} for one--particle potentials. It is
easy to see that they are exact already in first order in the one--body interaction.
For completeness we give $\mat{\Sigma}(\omega)$ and  $\mS(\omega)$
explicitly in these cases:
\beqs
	{\Sigma_{pq}}(\omega) &=& h_{pq}\\
	\Self_{rs,r's'}(\omega) &=& h_{rr'}\,\delta_{ss'} + h_{ss'}\,\delta_{rr'} -
	h_{rs'}\,\delta_{sr'} - h_{sr'}\,\delta_{rs'}
\eeqs
These matrices can be seen as representations of the one--particle Hamiltonian in the spaces
spanned by one--particle wave functions $a^\dagger_q\kv$ and two--particle Slater determinants
$a^\dagger_{r'} a^\dagger_{s'} \kv$, respectively.

Speaking in terms of resolvents, we see that both our extended two--particle propagator as well as
the \opg\ now become resolvents of energy independent effective Hamiltonians and these Hamiltonians
contain no higher  than first order contributions in the interaction.
We want to stress that this is {\em not} the case for the \papa propagator $\mat{\Pi}(\omega)$
from (\ref{def_papaprop})
or any of its constituents $\mat{\Pi}^\pm(\omega)$.
We already pointed out that these propagators can be viewed as resolvents only in restricted index 
spaces. Therefore, the behaviour of their matrix inverses under transformations of the orbital
basis is complicated and leads to {\em energy--dependent} effective Hamiltonians with contributions
in all orders of perturbation theory. For this reason the calculation of scattering amplitudes or 
double ionization energies via the \papa propagator or its constituents in the presence of
one--particle potentials is formally nearly as complicated as the treatment of a fully correlated
system (as long as one does not transform to new orbitals, of course). 
When using our new extended two--particle propagators, however, already the first
order contribution to its self energy yields the influence of the one--particle potentials
exactly.

\subsection{Time Space Equations}

For the sake of conceptual clearness we have developed the algebra of the extended two--particle
propagators in energy space. In order to apply our developments to the scattering problem of
section \ref{scat_sect}, we will consider briefly the Fourier transforms into time space.
We use the following convention for Fourier transformation:
\beq
	{f}(t) := \int_{-\infty}^\infty \frac{d\omega}{2\pi} e^{-i\omega t}f(\omega)
\eeq
In order to agree with standard notation we define:
\beqs
	\nonumber
	\cG_{rs,r's'}(t,t') &:=& {\cG}_{rs,r's'}(t-t') \\
	\nonumber
	\Pi_{rs,r's'}(t,t') &:=& {\Pi}_{rs,r's'}(t-t')
\eeqs
When transforming the extended two particle propagator $\cG_{rs,r's'}(\omega)$, it is
most convenient to start from Eq.~(\ref{cG_ast}) and make use of the convolution theorem.
The result reads:
\beqs \label{timespacecG}
	\nonumber
	\cG_{rs,r's'}(t,t') &=& \Pi_{rs,r's'}(t,t') \\
	\nonumber
	& & + \theta(t'-t) \left\{ \frac{1}{i}\bv \,a_s(t) a_{s'}^\dagger(t') \, \kv
	\cdot  \bpsi\, a^\dagger_{r'}(t') a_r(t)\, \kpsi\right\} \\
	& &- \{ r \longleftrightarrow s \}- \{ r' \longleftrightarrow s' \}
	+ \{ r \longleftrightarrow s\,,\,r' \longleftrightarrow s' \}
\eeqs
where the Fourier transform of the \papa propagator from Eq.~(\ref{def_papaprop})
is given by:
\beqs \label{timespacePi}
	\nonumber
	\Pi_{rs,r's'}(t,t') &=&  \frac{1}{i} \theta(t-t')\bpsi\,a_r(t) a_s(t) 
		a_{s'}^\dagger(t') a^\dagger_{r'}(t')\, \kpsi  \\
	& &  - \frac{1}{i} \theta(t'-t) \bpsi\,a_{s'}^\dagger(t') a^\dagger_{r'}(t') 
		  a_r(t) a_s(t)\, \kpsi
\eeqs

The two terms on the right hand side of the above equation represent the parts $\Pi^+_{rs,r's'}(t,t')$ 
and $\Pi^-_{rs,r's'}(t,t')$, respectively.
In these expressions the operators  appear in the Heisenberg picture with a time dependence
given by
\beq \label{defHeisenbergpicture}
	\hat{A}(t) = e^{i\hat{H}t}\hat{A} e^{-i\hat{H}t}.
\eeq
Heavyside's theta function is defined as follows:
\beq
	\theta(t)  := \left\{\begin{array}{l} 1 \quad\mbox{for $t\geq 0$}\\ 
						0 \quad\mbox{for $t < 0$}
	\end{array} \right.
\eeq
As usual, the signs of the infinitesimals $\pm i\eta$ bring about the time ordering
via theta functions. Although the time ordering does have relevance in scattering
by giving the right boundary conditions, it does not affect the algebraic properties,
since Dyson's equation transforms, regardless of the chosen time ordering of $\mcG$,
into
\beqs \label{dyseqtd}
	\nonumber \lefteqn{
	\cG_{rs,r's'}(t,t') = \cG^\zero_{rs,r's'}(t,t')} \\
	& & + \sum_{pq \atop p'q'} \int_{-\infty}^\infty
	dt_1\int_{-\infty}^\infty
	 dt_2\, \cG^\zero_{rs,pq}(t,t_1)\Self_{pq,p'q'}(t_1,t_2) \cG_{p'q',r's'}(t_2,t')
\eeqs
The zeroth order propagator  in time space reads:
\beqs \label{Gzerotimed}
	\nonumber
	\cG^\zero_{rs,r's'}(t,t') &=& (\ddrs) \frac{1}{i} e^{i(\varepsilon_r + \varepsilon_s)(t-t')}\\
	& &\cdot
	[\overline{n}_r\overline{n}_s\theta(t-t') - (1-\overline{n}_r\overline{n}_s)\theta(t'-t)]
\eeqs

\section{Optical Model Description of Two--Particle Scattering} \label{s2}
 	\label{scat_sect}

In this chapter we will show how an extended two--particle propagator that satisfies Dyson's equation
can be used to derive effective two--particle equations for the elastic scattering of a two--fermion
projectile by a correlated many--body target.\footnotemark
\ The extended propagator introduced in the preceding chapter serves as one example, but other
two particle Green's functions can be used as well, provided they fulfil certain requirements
which will be discussed.
We will show how the improper self energy of this propagator can be identified with the scattering
T--matrix. Furthermore, we will derive an effective Lippmann--Schwinger equation and an eigenvalue
equation involving a two--particle scattering amplitude and an optical potential that is given by the 
two--particle self energy.

\subsection{The Scattering Problem} \label{NotAPreq}

The type of scattering process that can be straightforwardly described by the considered type of
two particle propagators is the elastic scattering of a two particle  projectile off a correlated
many--body target. For simplicity we assume that all particles are indistinguishable fermions,
although extensions to composite particle systems, e.~g.~a positronium projectile, are easy to do
and usually simplify the actual calculations. The target is taken to be in its nondegenerate
$N$--fermion ground state $\kpsi$ with energy $\Eo$.
In the simplest case, the projectile may consist of two (asymptotically) free fermions and is denoted
by:
\[
	a_\p^\dagger a_\q^\dagger\,\kv
\]
It is an eigenstate of the `free particle' Hamiltonian which can be identified with the
zeroth order Hamiltonian $\Ho$ from Eq.~(\ref{defHo}). All relevant information
on the scattering process is now contained in the two particle part $\mat{\Pi}^+(\omega)$ 
of the \papa propagator from Eq.~(\ref{def_papa_plus}).
However, we may consider a more general projectile consisting of an arbitrary
two particle eigenstate $\ket{\K}$ of some ``free projectile'' Hamiltonian $\Hp$.
\beqs
	\Hp\,\ket{\K} &=& \E_\K\,\ket{\K} \\ \label{defK}
	\ket{\K} &:=& A_\K^\dagger\,\kv 
\eeqs
with $A_\K^\dagger$ denoting the projectile creation operator:
\beqs \label{defAK}
	A_\K^\dagger &=& \int d\p\,d\q\,\phi_\K(\p,\q)\, a_\p^\dagger a_\q^\dagger
\eeqs
Here the cumulative index $\K$ denotes a complete set of quantum numbers that characterize the two particle
state $\ket{\K}$ with energy $\E_\K$. 
Let $\p$ and $\q$ denote one--particle quantum numbers, e.~g.~momentum 
and spin $\p = (\vec{p},\sigma)$. 
Since it appears more natural to use continuous sets of quantum numbers rather than discrete
orbital indices when describing scattering processes, we changed the notation and replaced
sums over one or two particle indices by integrals. This is merely a formal step, of course, since
the integrals still contain discrete sums, e.~g.~over spin indices.
The operators $a_\p^\dagger$ and $a_\q^\dagger$ create one particle
states with the given quantum numbers. The quantity $\phi_\K(\p,\q)$ plays the role of a
two particle wavefunction.

The simple case of a projectile consisting of two free particles now emerges as a special case in
the more general treatment. The free projectile Hamiltonian $\Hp$ reduces to 
the kinetic energy operator and we can take  $a_\p^\dagger$ and $a_\q^\dagger$ to create plane wave states.
In that case we can choose the wavefunction to be of the form
\beq
	\phi_\K(\p,\q) = \delta_{\ik_1 \p}\,\delta_{\ik_2 \q}
\eeq
where $\K\equiv(\ik_1,\ik_2)$.
In the following we will develop the theory for a general projectile state, but come back to the simpler
case at times.
 
The principal quantity for describing the results of scattering experiments is the so called {\em scattering matrix}
or {\em S--matrix:}
\beq \label{defscat}
	\scat{\K}{\K'} = \bracket{\Psi^-_\K}{\Psi^+_{\K'}}
\eeq
We want to describe the elastic scattering of two particle states $\ket{\K}$.
The S--matrix then gives the probability amplitude for the scattering of the two particle projectile $\ket{\K'}$
off the target state $\kpsi$, into the projectile state $\ket{\K}$, the target remaining in the state $\kpsi$.
Formula (\ref{defscat}) gives the S--matrix as the overlap of the so called {\em scattering states:}
\beq \label{scatstate}
	\ket{\Psi^\pm_\K} = \lim_{t\rightarrow \mp \infty}A_\K^\dagger(t)\kpsi e^{-i\E_\K t}
\eeq
Here the projectile creation operator $A_\K^\dagger$ appears in the Heisenberg picture defined in
Eq.~(\ref{defHeisenbergpicture}). The scattering states can be understood as states in 
the Heisenberg picture 
that behave asymptotically in the distant past/future
like a free projectile $\ket{\K}$ and target $\kpsi$.

Strictly speaking, the limit in (\ref{scatstate}) does not 
converge in the usual sense and is meant as an Abelian
limit\cite{Simon}:
\beq
	\ket{\Psi^\pm_\K} = \lim_{\varepsilon\searrow 0}\int_0^{\mp\infty}d\tau\,e^{\pm\varepsilon\tau}
	\,A_\K^\dagger(\tau)\kpsi e^{-i\E_\K \tau}
\eeq
In the following we will use the Abelian limit where appropriate, without referring explicitly to it.

For the S--matrix we find:
\beq \label{Sexpectation1}
	\scat{\K}{\K'} = \lim_{t\rightarrow \infty \atop t'\rightarrow - \infty} e^{i\E_\K t}\bpsi\, A_\K(t)\,
	 A_{\K'}^\dagger (t')\,\kpsi e^{-i\E_{\K'} t'}
\eeq
Employing the definition (\ref{defAK}) of the projectile creation operator $A_{\K}^\dagger$, we arrive at the
following expression:
\beqs \label{Sexpectation}
	\nonumber \lefteqn{
	\scat{\K}{\K'}  = \lim_{t\rightarrow \infty \atop t'\rightarrow - \infty} \,e^{i(\E_\K t - \E_{\K'} t')}
	\int d\p\,d\q\,d\p'\,d\q'\, \cdot }\\
	& & \cdot\, \phi^*_\K(\p,\q)\,\bpsi\, a_\q(t)\, a_\p(t)\,
	 a_{\p'}^\dagger(t')\, a_{\q'}^\dagger(t')\,\kpsi\,\phi_{\K'}(\p',\q') \\
	\nonumber
	&=& \lim_{t\rightarrow \infty \atop t'\rightarrow - \infty} \,e^{i(\E_\K t - \E_{\K'} t')}
	\int d\p\,d\q\,d\p'\,d\q'\, \cdot \\
	& & \cdot\, \phi^*_\K(\p,\q)\,{\Pi}^+_{pq,p'q'}(t,t')\,\phi_{\K'}(\p',\q')
\eeqs
The last equality follows the definition (\ref{timespacePi}). This shows that the two particle part 
$\mat{\Pi}^+(t,t')$ of the \papa propagator contains all relevant information on the scattering process.
Obviously, $\mat{\Pi}^+(t,t')$ can be replaced by $\mat{\cG}(t,t')$ as well as by $\mat{\Pi}(t,t')$,
since the additional terms do not contribute
due to the fixed time ordering in equation(\ref{Sexpectation}), like it can be seen from 
equations (\ref{timespacecG}) and (\ref{timespacePi}).

\subsection{General requirements for a scattering propagator}\label{sectreqi}

The main objective of this chapter is to show how the extended two particle propagator $\mat{\cG}$ of
Eq.~(\ref{def_cG}) or (\ref{timespacecG}) can be used to reduce the many--body scattering problem to
effective two--particle scattering equations.
The derivation we will give in the following, however, 
only makes use of some fundamental properties of the extended propagator.
Loosely speaking these are: the propagator used contains $\mat{\Pi}^+$ and satisfies Dyson's equation.
Since it is possible
to find different extensions of the \papa propagator that meet these requirements, we will introduce
the new symbol $\cGs_{\K,\K'}(t,t')$ for a general propagator that fulfils certain requirements
stated below.
We will refer to the general two particle Green's function $\cGs_{\K,\K'}(t,t')$ as the 
{\em scattering propagator}. For convenience this propagator is labelled with two particle
quantum numbers like $\K$ and $\K'$, each of these replacing  two orbital indices.
The transformation between these two methods of labelling two particle propagators is given
by the two particle wavefunctions $\phi_\K(\p,\q)$ like in Eq.~(\ref{defAK}).
In the following we will state the {\em required properties} explicitly:

The first requirement allows one to rewrite Eq.~(\ref{Sexpectation1}) for the S--matrix in terms
of the scattering propagator:
\begin{itemize}
\item[(i)]	For $t>t'$ the scattering propagator reduces to the transformed  two 
		particle propagator $\mat{\Pi}^+(t,t')$:
		\beq \label{itemi}
			i\,\cGs_{\K,\K'}(t,t') =  \bpsi\, A_\K(t)\,A_{\K'}^\dagger(t')\,\kpsi
			\qquad\mbox{for $t>t'$}
		\eeq
\end{itemize}
The second property requires the main new feature of our extended two particle propagator:
Dyson's equation. It introduces the self energy $\Selfs$ of the scattering propagator $\cGs$,
which will serve as the two particle optical potential:
\begin{itemize}
\item[(ii)]	The scattering propagator satisfies a Dyson equation:
		\beqs \label{dyseq1}
			\nonumber\lefteqn{
			\cGs_{\K,\K'}(t,t') = \cGs^\projectile_{\K,\K'}(t,t')} \\ & &+
			\int d\K_1\,d\K_2\,dt_1\,dt_2\, \cGs^\projectile_{\K,\K_1}(t,t_1)\,
			\Selfs_{\K_1,\K_2}(t_1,t_2)\,\cGs_{\K_2,\K'}(t_2,t')
		\eeqs
		Here, a {\em free projectile} propagator $\cGs^\projectile$ appears that reads
		\beq \label{freecond} \label{addfreecond}
			i\,\cGs^\projectile_{\K,\K'}(t,t') = \theta(t-t')\,\delta_{\K,\K'}\, e^{i\E_\K(t-t')}
		\eeq
		for $\K$ or $\K'$ belonging to the subset of two particle quantum numbers
		used to describe the projectile states.
\end{itemize}
The integrals over $\K_i$ in (\ref{dyseq1}) are meant to run over the complete two particle space. The 
$\delta$ function 
denotes a proper normalization condition. The actual
form of this condition depends on the set of quantum numbers used.
In the example of two free projectile particles this normalization condition reads:
\beq
	\delta_{\K,\K'} = \delta_{\ik_1 \ik_1'}\,\delta_{\ik_2 \ik_2'}-
	\delta_{\ik_1 \ik_2'}\,\delta_{\ik_2 \ik_1'}
\eeq
The second term on the right hand side accounts for the fermionic nature of the projectile's
components.
In the case of a bound state projectile, on the  other hand, the index $\K$ may combine an internal
quantum number $n$ of the projectile with a center of mass wavevector $\vec{k}$ and consequently
the $\delta$ function is given by:
\beq
	\delta_{\K,\K'} = \delta_{n n'}\,\delta_{\vec{k} \vec{k}'}
\eeq

The third requirement is a minor technical point. It is a common feature of many--body Green's
functions depending on two time variables that these functions actually depend
only on the time difference, when the total energy of the given system is conserved.
Thus we require:
\begin{itemize}
\item[(iii)]	As a consequence of the conservation of energy we want the function $\cGs$ and the other
		functions that appear in the Dyson equation to depend only on the
		time difference $t-t'$.
\end{itemize}

It is easy to see that these conditions are fulfilled by the extended propagator 
$\cG_{\ik_1 \ik_2, \ik_1'\ik_2'}(t, t')$ of Eq.~(\ref{timespacecG}), in the case of a projectile 
consisting of two (asymptotically) free
particles where $\K\equiv(\ik_1,\ik_2)$ holds. The zeroth order propagator 
$\cG^\zero_{\ik_1 \ik_2, \ik_1'\ik_2'}(t, t')$ from Eq.~(\ref{Gzerotimed}) now can be identified with
the free projectile propagator $\cGs^\projectile_{\K,\K'}(t,t')$.

In the more general case of a projectile in an arbitrary, possibly bound, two--particle state, 
the transformed function
\beq
	\cGs_{\K,\K'}(t,t') :=\int d\p\,d\q\,d\p'\,d\q'\, \phi^*_\K(\p,\q)\,\cG_{pq,p'q'}(t, t')\,\phi_{\K'}(\p',\q')
\eeq
fulfils the requirements. We only have to be a little more careful,
if the projectile state $\ket{\K}$ is not a Slater determinant and thus cannot be expressed as the eigenstate
of a one--particle operator. This is generally true for bound two--particle projectile states.
Since the zeroth order Hamiltonian  $\Ho$ of Eq.~(\ref{defHo}) is usually required to be a one--particle operator,
the free projectile propagator $\cGs^\projectile$ of Eq.~(\ref{freecond}) will generally differ from
the zeroth order $\cG^\zero$ of Eq.~(\ref{Gzerotimed}) both in the phases and the set of quantum numbers that
diagonalizes the propagator.
This does not cause a serious problem, however, since the Dyson equation (\ref{dyseqtd}) for $\cG$ can be rewritten with
$\cG^\zero$  replaced by the propagator $\cGs^\projectile$.
This implies a trivial renormalization of the $\omega$ {\em independent} part of
the self energy $\mat{\Self}(\omega)$, as the energy--space equations  (\ref{dys2}) and (\ref{dys3}) show.

	If we still want to keep the direct connection to perturbation theory, such that
	$\cGs^\projectile$ is the zeroth order of $\cGs$, we have to find some way to include $\Hp$,
	which is  primarily an operator in the two particle Hilbert space, into the zeroth order
	(Fock--space) Hamiltonian $\Ho$. In this case
	we have to take care that the free projectile states do not interfere with the 
	zeroth order ground state of the target, which also has to be an eigenstate of $\Ho$.
	This could be done by constructing the projectile from unoccupied (virtual) orbitals only.
	A procedure similar to this is described in reference \cite{junkin_I}.

We now carry on without referring to the extended propagator introduced in the last chapter but assume
that we have a Green's function, the scattering propagator $\cGs$, which possesses the properties stated above.

\subsection{Relation to the Scattering T--Matrix}

Now we are in the position to link up with Eq.~(\ref{Sexpectation1}) and 
express the S--matrix in terms of the scattering propagator using property {{(i)}}:
\beq \label{scatGreen}
	\scat{\K}{\K'} = \lim_{t\rightarrow \infty \atop t'\rightarrow - \infty} i\,\cGs_{\K,\K'}(t,t') 
	\,e^{i(\E_\K t-\E_{\K'} t')}
\eeq
In order to see the connection to the popular T--matrix description of scattering we rewrite Dyson's equation
(\ref{dyseq1}) 
introducing the {\em improper self energy} or Green's function {T--matrix} $\cTs_{\K,\K'}(t,t')$:
\beq \label{dysT}
  \begin{array}{rcl}
	\mat{\cGs} &=& \mat{\cGs}^\projectile + \mat{\cGs}^\projectile\mat{\cTs}\,\mat{\cGs}^\projectile \vspace{\jot} \\
	\mat{\cTs} &=& \mat{\Selfs} + \mat{\Selfs}\,\mat{\cGs}^\projectile\mat{\cTs}
  \end{array}
\eeq
Here, we are using a shorthand notation where the entities like $\mat{\cGs}$ are matrices with indices
$(\K,t)$ and $(\K',t')$. The matrix product then implies integration over all inner indices, namely the two
particle quantum numbers and the time variable. For example:
\beqs
	\nonumber \lefteqn{
	\left[\mat{\cGs}^\projectile\mat{\cTs}\,\mat{\cGs}^\projectile\right]_{(\K,t)(\K',t')} =} \\
	& &  \int d\K_1\,d\K_2\,dt_1\,dt_2\,
	\cGs^\projectile_{\K,\K_1}(t,t_1)\,\cTs_{\K_1,\K_2}(t_1,t_2)\,\cGs^\projectile_{\K_2,\K'}(t_2,t')
\eeqs

It is easily found that inserting (\ref{dysT}) into the expression (\ref{scatGreen})  
for the S--matrix and using property {(i)} yields:
\beq \label{scatT}
	\scat{\K}{\K'} =\delta_{\K \K'} -2\pi i\,\cT_{\K,\K'}(\E_{\K})\,\delta(\E_{\K}-\E_{\K'})
\eeq
Here, the quantity $\cT_{\K,\K'}(\E_{\K})$ denotes the Fourier transform of the improper self energy
$\cT_{\K,\K'}(t,t')$, evaluated at energy $\E_{\K}$. Transforming $\mat{\cT}$ on a single energy variable
becomes possible, because the improper self energy is a
function of the time difference only. This is a consequence of property {(iii)}.

Eq.~(\ref{scatT}) now allows us to identify the improper self energy $\cT_{\K,\K'}(\E_{\K})$
with the {\em on--shell scattering T--matrix} $\scatT{\K}{\K'}$, which is defined by
\beq
	\scat{\K}{\K'} =\delta_{\K \K'} -2\pi i\,\delta(\E_{\K}-\E_{\K'})\,\scatT{\K}{\K'}.
\eeq

This shows a
connection between the many--body Green's function $\cGs$ and effective scattering quantities represented by 
the common scattering S-- and T--matrices, which
is analogous to the problem of scattering one fermion off a correlated many fermion
target \cite{RingSchuck,CzsanakTY}. In that case the \opg\ $\mat{G}(t,t')$ takes on the
role of $\mat{\cGs}(t,t')$.

\subsection{Effective Lippmann--Schwinger Equation}

We will now proceed to further carry this analogy and derive effective two--particle scattering equations.
First we define {\em scattering amplitudes} that take the role of effective two--particle wavefunctions:
\beq \label{ampl}
	f^{\K' +}(\K,t) := \bpsi A_{\K}(t) \ket{\Psi^+_{\K'}}
\eeq
These amplitudes contain all the relevant information concerning the scattering process, since 
we can express the S--matrix through:
\beq \label{Sthroughf}
	\scat{\K}{\K'} = \lim_{t\rightarrow \infty} e^{i\E_\K t}f^{\K' +}(\K,t) 
\eeq
The relation to the Green's function is given by:
\beq \label{amplG}
	f^{\K' +}(\K,t) = \lim_{t'\rightarrow -\infty} i\,\cGs_{\K,\K'}(t,t') 
	e^{-i\E_{\K'} t'}
\eeq

The time dependence of the scattering amplitudes can be taken from the definition, using the fact that
$\kpsi$ and $\ket{\Psi^+_{\K'}}$ are eigenfunctions of the Hamiltonian:
\beqs
	\nonumber
	f^{\K' +}(\K,t) &=&  \bpsi A_{\K}(t) \ket{\Psi^+_{\K'}} \\
	\nonumber
	&=& \bpsi e^{i \hat{H}t} A_{\K} e^{-i \hat{H}t} \ket{\Psi^+_{\K'}} \\
	&= & e^{-i\E_{\K'} t} f^{\K' +}(\K)
\eeqs
The last equality defines the {\em time independent scattering amplitude}. We note that
\beqs \label{tiAmpl}
	f^{\K' +}(\K) &=& e^{i\E_{\K'} t} f^{\K' +}(\K,t) \\ 
	&=& \bpsi  A_{\K} \ket{\Psi^+_{\K'}}
\eeqs

We can now use Dyson's equation for  $\cGs$ to derive an effective Lippmann--Schwinger equation
for the time independent scattering amplitude $f^{\K' +}(\K)$. Inserting Dyson's equation (\ref{dyseq1})
into  (\ref{amplG}) and using  (\ref{tiAmpl}) leads after some simple manipulations to the equations:
\beqs
	\nonumber
	f^{\K' +}(\K) &=& \lim_{t'\rightarrow -\infty} i\,e^{i\E_{\K'} t}\,\cGs_{\K,\K'}(t,t')\, 
	e^{-i\E_{\K'} t'} \\
	\nonumber
	&=& \delta_{\K \K'} +  \lim_{t'\rightarrow -\infty} i \int  d\K_1\,d\K_2\,dt_1\,dt_2\, \times \\
	& & e^{i\E_{\K'} t}\,\cGs^\projectile_{\K,\K_1}(t,t_1)\,\Selfs_{\K_1,\K_2}(t_1,t_2)\,\cGs_{\K_2,\K'}(t_2,t')
	\,e^{-i\E_{\K'} t'}
\eeqs
Carrying out the time integrations with the help of property {(iii)} leads to the desired working equation
for the time independent scattering amplitude:
\beq \label{scateqn}
	f^{\K' +}(\K) = \delta_{\K \K'} + \int  d\K_1\,d\K_2\,\cGs^\projectile_{\K,\K_1}(\E_{\K'})
	\,\Selfs_{\K_1,\K_2}(\E_{\K'})\, f^{\K' +}(\K_2)
\eeq
This is an integral equation which can be solved for  the scattering amplitude once the self 
energy $\mat{\Selfs}(\omega)$ is known. It is  analogous to the Lippmann--Schwinger equation of basic scattering
theory. In the above equation (\ref{scateqn}) the self energy replaces the potential that appears in the
Lippmann--Schwinger equation. Therefore we have actually derived an exact optical model description for 
our many--body 
scattering problem: The initial $N+2$ particle problem has been reduced to the problem of scattering two
particles by an optical potential. The optical potential, which is given by the self energy $\mat{\Selfs}(\omega)$,
is in general energy dependent and takes account of 
all the correlation within the target as well as the interaction
between the target and the projectile particles. Like we have seen in section \ref{sectOneP},
this self energy becomes energy independent for systems with particles interacting
by one particle potentials only.

All this is analogous to  the elastic scattering of one fermion by a many--body target.
In this case the self energy $\mat{\Sigma}(\omega)$ belonging to the \opg\ $\mat{G}(\omega)$ takes the
role of the optical potential\cite{RingSchuck,CzsanakTY,bellandsquires,namiki}, which becomes energy independent
for systems interacting by a one particle potential.

\subsection{A Two--Particle Eigenvalue Equation} \label{eigenv_eqn}

In basic scattering theory the Lippmann--Schwinger integral equation presents an alternative formulation
to Schr\"odinger's eigenvalue equation. Thus we can expect that in an optical model description
it is, in general, possible to formulate an effective eigenvalue equation for the
scattering amplitudes.
Multiplying the integral equation (\ref{scateqn}) with the 
energy difference $(\E_{\K'} -\E_{\K})$ and using the Fourier transform of 
(\ref{addfreecond}) leads us
to the desired equation:
\beq \label{eigenveqn}
	(\E_{\K} -\E_{\K'})\,f^{\K' +}(\K) + \int  d\K_2\,\Selfs_{\K,\K_2}(\E_{\K'})\, f^{\K' +}(\K_2) = 0
\eeq
This is an eigenvalue equation in the two particle Hilbert space. It is formulated using a basis of two particle
wavefunctions that diagonalize the projectile Hamiltonian $\Hp$ defined in \ref{NotAPreq}. Reintroducing
$\Hp$ and switching from the quantum number representation to abstract two--particle--space vectors
we can rewrite (\ref{scateqn}) into:
\beq \label{eigenvEqn}
	\left\{\Hp + \hat{\Self}(\E_{\K'})\right\}\,\Ket{f^{\K'}+} = \E_{\K'}\,\Ket{f^{\K'}+}
\eeq
We have defined the effective scattering kets $\ket{f^{\K'}+}$ with the help of the free projectile 
states $\ket{\K}$ of Eq.~(\ref{defK}) by
\beq
	\bracket{\K}{f^{\K'}+} = f^{\K' +}(\K)
\eeq
The operator $\hat{\Self}(\E_{\K'})$ is defined by it's matrix elements:
\beq
	\bra{\K_1}\hat{\Self}(\omega)\ket{\K_2} = \Selfs_{\K_1,\K_2}(\omega)
\eeq
Of course, the Green's function $\cGs_{K_1,K_2}(\omega)$ is now rediscovered as a representation
of a resolvent $\hat{\cG}(\omega)$ that satisfies the following operator identity in two particle 
Hilbert space:
\beq \label{effhamid}
	\left\{\Hp + \hat{\Self}(\omega) - \omega\right\}\, \hat{\cG}(\omega) \equiv \hat{\id}
\eeq
This links up with section \ref{sectResolvents}, where we already interpreted the extended
two  particle propagator as a resolvent  of an $\omega$ dependent operator. Now we have finally
proven that the scattering propagator (and therefore also the extended two particle propagator from the last
chapter) actually is the resolvent of an  effective Hamiltonian 
that allows to calculate S--matrix elements through (\ref{eigenvEqn}) and (\ref{Sthroughf})!
Naturally, the remarks of section  \ref{sectResolvents} on the energy dependence of the
effective Hamiltonian apply here in the same way. 

We want to remind of the fact that the effective Hamiltonian and therefore also the
optical potential derived directly by Feshbach projection contains a divergence that is avoided
by our two particle self energy. Also the behaviour with respect to one particle potentials
discussed in section \ref{sectOneP} provides a contrast between the two approaches.
This leads us to the following conclusion:
Given some arbitrary two particle propagator,  it is the properties (i), (ii), and (iii) of 
section \ref{sectreqi} that lead to 
the effective scattering equations of this last chapter. The usefulness of these equations for
approximative calculations of scattering parameters, however, is determined by the properties of
its self energy. As shown in chapter \ref{s3}, the self energy related to our extended propagator
possesses a good--natured energy dependence and behaves well with respect to one particle potentials.
For these reasons  the extended two particle propagator provides a useful tool in scattering.
Furthermore it presents an alternative approach that removes drawbacks of the Feshbach projection effective
potentials discussed in reference \cite{junkin_I}.

\section{Conclusions} \label{conclusion}

In this paper we have presented an extension of the fermionic \papa propagator that has algebraic
properties similar to the  \opg. We explicitly constructed one example of an extended propagator and 
investigated its major algebraic properties. These include an analogon of Dyson's equation with a 
two--particle self energy. We showed that this self energy serves as an  exact optical potential for 
the scattering of a two--particle projectile by a correlated many--body target. The projectile may
either consist of two asymptotically free fermions or a  bound two--particle state. In the latter
case the two--particle self energy has to be trivially renormalized.

In contrast to effective two particle potentials derived directly from the  multiple particle
Schr\"odinger equation, the two particle self energy has a good--natured energy dependence.
Additionally,  correlations due to a one--particle potential are accounted for exactly already by a first
order, energy--independent  self energy.

The ideas behind the construction of extended two--particle propagators are quite general and lead to a 
whole new class of extended two--particle propagators with similar algebraic properties. Employing 
approximations that aim at ab initio calculations will lead to different working equations for each
of these functions.
A new feature of the extended propagators is the  appearance of degeneracies that may require the
use of degenerate techniques necessary when employing perturbation theory based approximations.
The reason for this lies in the indefiniteness of the metric used in the construction of the 
extended propagators. It is an open question, whether or not an extended propagator can
be found that combines
a definite metric with the advantageous properties we discussed in the present paper.

The present approach can be generalized to construct optical potentials for the scattering of an
$M$--particle projectile by a many--body target. The starting point is the introduction of an extended
$M$--particle propagator which fulfills a Dyson equation.




\section*{Notes}

\begin{enumerate}

\item {\label{regeln}
	We use the following definition for the scalar product, which is canonical:
  \[
	\left\langle\dv{\ket{\psi_1}}{\bra{\psi_2}}, \dv{\ket{\phi_1}}{\bra{\phi_2}} \right\rangle
	:=\bracket{\psi_1}{\phi_1} + \bracket{\phi_2}{\psi_2}
  \]
	The action of a diagonal operator is defined by:
  \[
	  \left(\begin{array}{cc}
	\hat{A}_1	&	0	\\
	0	&	\hat{A}_2	\\ \end{array} \right)\dv{\ket{\phi_1}}{\bra{\phi_2}} 
	:=\dv{\hat{A}_1\,\ket{\phi_1}}{\bra{\phi_2}\,\hat{A}_2} 
  \]
}

\item {
	Instead of the matrix partitioning carried out in \cite{alessandra}, 
	one can equally well proceed by introducing
	projection operators onto the subspace spanned by the set of vectors $\{\ket{Y_q^\G}\}$ and 
	its orthogonal complement\cite{CapuzziMahaux}.}  

\item {
For these states the rules of footnote \ref{regeln} apply likewise. Additionally we define
for product states:
\begin{eqnarray*}
	\big\langle\ket{a'}\otimes\bra{b'},
	\ket{a}\otimes\bra{b}\big\rangle &=& \bracket{a'}{a}\cdot\bracket{b}{b'} \\
	\mbox{and}\qquad\qquad (\hat{A}\otimes\hat{B})\,
	(\ket{a}\otimes\bra{b}) &=& \hat{A}\ket{a}\otimes\bra{b}\hat{B} 
\end{eqnarray*}
}

\item {For earlier work on effective two--particle
scattering equations in a different context see \cite{junkin_I}.}

\end{enumerate}



\end{document}